\documentclass[12pt,draftclsnofoot,onecolumn]{IEEEtran}
\usepackage{amssymb}
\usepackage{amsmath}
\usepackage{amsfonts}
\usepackage{graphicx}
\usepackage{subfigure}
\usepackage{cite}
\usepackage{enumerate}
\usepackage{url}
\usepackage[ruled]{algorithm2e}
\usepackage{algorithmicx}
\usepackage{mathtools}
\usepackage{makecell}
\allowdisplaybreaks[4]

\begin{document}

\newtheorem{definition}{\bf Definition}	
	
\title{\huge{Hypergraph Theory: Applications in 5G Heterogeneous Ultra-Dense Networks}}
\author{
\IEEEauthorblockN{
\normalsize{Hongliang Zhang}\IEEEauthorrefmark{1},
\normalsize{Lingyang Song}\IEEEauthorrefmark{1},
\normalsize{Yonghui Li}\IEEEauthorrefmark{2},
\normalsize{and Geoffrey Ye Li}\IEEEauthorrefmark{3}\\}
\IEEEauthorblockA{\small
\IEEEauthorrefmark{1}School of Electrical Engineering and Computer Science, Peking University, Beijing, China, \\Email: $\left\{\mbox{hongliang.zhang, lingyang.song} \right\}$@pku.edu.cn.\\
\IEEEauthorrefmark{2} School of Electrical and Information Engineering, The University of Sydney, Sydney, Australia, \\Email: yonghui.li@sydney.edu.au.\\
\IEEEauthorrefmark{3}School of Electrical and Computer Engineering, Georgia Institute of Technology, Atlanta, GA, USA \\Email: liye@ece.gatech.edu.}

}

\maketitle
\begin{abstract}
\normalsize{ Heterogeneous ultra-dense network (HUDN) can significantly increase the spectral efficiency of cellular networks and cater for the explosive growth of data traffic in the fifth-generation~(5G) communications. Due to the dense deployment of small cells~(SCs), interference among neighboring cells becomes severe. As a result, the effective resource allocation and user association algorithms are essential to minimize inter-cell interference and optimize network performance. However, optimizing network resources in HUDN is extremely complicated as resource allocation and user association are coupled. Therefore, HUDN requires low-complexity but effective resource allocation schemes to address these issues. Hypergraph theory has been recognized as a useful mathematical tool to model the complex relations among multiple entities. In this article, we show how the hypergraph models can be used to effectively tackle resource allocation problems in HUDN. We also discuss several potential research issues in this field.}
\end{abstract}

\section{Introduction}%
In recent years, the rapid growth of various booming wireless communication services has led to an explosion of wireless data traffic. A 10-fold increase in the mobile data from 2014 to 2019 has been predicted in~\cite{Cisco-2015}, and this trend may accelerate in the future. As a result, the exponential traffic growth requires an ever higher capacity for the future fifth-generation~(5G) wireless networks. To accommodate such a significant growing demand, network densification technology~\cite{NJDRDARCS-2014} has emerged as a promising solution to significantly increase the network spectral efficiency through dense deployment of small cells~(SCs) in the heterogeneous cellular network, that is, the network consisting of multiple radio access technologies, instead of a single type. As such, the small cell access points~(SAPs) can get as close as possible to the end users, thus leading to efficient spatial reuse of network resources.

In general, heterogeneous ultra-dense networks~(HUDNs) have two key features:
\begin{itemize}
	\item \textbf{High SAP density:} In HUDN, the density of SAPs is much higher than that of active mobile users~\cite{MWA-2016}. The dense deployment of SCs will offload a significant amount of data traffic from macro cells to SCs.
	
	\item \textbf{Various types of SAPs:} The SAPs feature different capabilities, transmission powers, coverage, and deployment scenarios. The different operation modes of SAPs will influence the resource allocation.  
\end{itemize}

The ultra-dense deployment leads to some differences in resource allocation between HUDN and traditional networks, which can be summarized below: 

\begin{itemize}
	\item In traditional networks, the spectrum is reused in different cells. However, due to the dense deployment in HUDN, it is impractical to use the same spectrum in two proximate SCs.
	
	\item Interference among neighboring cells in HUDN is severer than that in traditional networks due to the close proximity of the SCs, and thus effective management is necessary.
	
	\item The association is another difference. In traditional networks, a user always connects to the base station~(BS) with the maximum signal to interference-plus-noise ratio~(SINR) while the HUDN considers a load-based association criterion~\cite{MWA-2016}. Besides, users in HUDN can connect to a macro cell and a SC simultaneously, referred to as the dual connectivity~\cite{CKHPSETB-2016}.
\end{itemize}

Since various types of SAPs are deployed with large density, many SCs can be interfered by a given user. Although the transmit power is low, the cumulative interference from multiple interfering sources, even if they are weak, may create a strong interference. In addition, resource allocation is also coupled with user association, which makes resource allocation extremely complicated in HUDN. Therefore, a low complexity but effective resource allocation technique is essential in order to explore its promising benefits in high spectral efficiency. 

Graph theory has proved to be a useful tool to solve resource allocation problems in wireless communications~\cite{HTLZ-2013}. However, the conception of the edge in graph theory can only model pairwise relations and is not sufficient in modeling cumulative interference in HUDN. Besides, graph is unable to model the relations among users, SAPs, or channels, either. In this regard, hypergraph theory, a more powerful mathematical tool, must be used. Several works have discussed the hypergraph application in resource allocation for wireless communications. Hypergraph has been used to capture accumulative interference accurately in~\cite{HLH-2016} and to model the complicated relations among content requesters, helpers, and channels in~\cite{LHYWH-2016}. Recently, we have found that hypergraph can be also applied in resource allocation problems.

In this article, we apply hypergraph approaches to model and analyze typical resource allocation problems in HUDN. Different densification schemes are taken into account, and the influence of user association is considered as well. Specifically, we provide four representative scenarios of HUDN and discuss their associations with different hypergraph models. 

Based on the structure of HUDN, we classify the network densification schemes into two categories: centralized and distributed network densification.

\begin{itemize}
	\item \textbf{Centralized network densification:} In a centralized HUDN, a central entity manages the network resource allocations. This centralized entity can be a BS equipped with a Distributed Antenna System~(DAS)~\cite{RSYJ-2013} or a pool of base band processing units~(BBUs) in a Cloud Radio Access Network~(CRAN)~\cite{AHYLGML-2014}. 
	
	\item \textbf{Distributed network densification:} Distributed scheme does not require any central entity for coordinating the channel access procedure. Due to lack of centralized coordination, this scheme requires scalable algorithms for the collaboration among different entities.
\end{itemize}

In the centralized network densification schemes, we consider CRAN and dual connectivity based densification scenarios, which can be modeled by hypergraph matching and hypergraph coloring, respectively. In the distributed network densification schemes, we consider device-to-device~(D2D) communications and proactive caching based densification scenarios, and apply hypergraph game as well as hypergraph matching to solve these resource allocation problems. 

The rest of this article is organized as follows. Section \ref{Scenario} presents four typical scenarios for resource allocation in HUDN. The corresponding hypergraph models are provided in Section~\ref{Resource}. A specific example of CRAN based densification is elaborated with more detailed evaluation in Section~\ref{OneExample}. In Section~\ref{Conclusion}, we conclude this article and identify possible research directions.

\section{Typical Scenarios in HUDN}
\label{Scenario}

In this section, we classify the resource allocation scenarios in HUDN into centralized and distributed network densification schemes. In each densification scheme, two typical scenarios and their resource allocation problems are introduced. 

\subsection{Centralized Network Densification}

For centralized network densification, we focus on two representative scenarios: CRAN based densification and dual connectivity based densification.
	
\textbf{CRAN based densification:} CRAN is an application for cloud computing in wireless networks, where the BBUs of BSs are pooled while the Remote Radio Heads~(RRHs) are used to provide the coverage, and they are connected by high capacity fibers. Pooling base band processing resources and coordinating signal processing can reduce the cost to deploy network, improve system performance, and simplify the implementations of multiple standards. 

\begin{figure}[!t]
	\centering
	\includegraphics[width=6.0in]{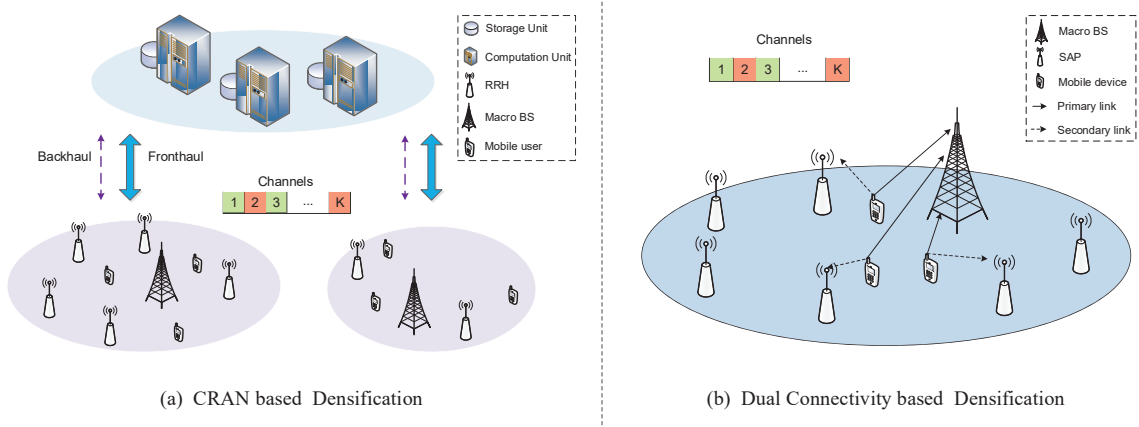}
	\caption{Two representative scenarios for centralized network densification.} \label{centralized}
\end{figure}

As shown in Fig.~\ref{centralized}(a), the CRAN based network densification is achieved via the deployment of numerous RRHs to carry the traffic from users. The communication process is briefly introduced in the following:
\begin{itemize}
\item RRHs act as an interface between the users and the BBU pool to perform the RF functions (e.g., filtering, power amplification, digital to analog conversion, and etc. \cite{AHYLGML-2014}). The user will connect to one RRH, and its data is conveyed to the BBU pool via the fronthaul link.

\item The computation and centralized control (e.g., channel allocation) are performed in the BBU pool. After computation, the result will be sent back to the mobile network via the backhaul links.
\end{itemize}

Hence, the resource allocation problem relates to the joint RRH association and channel access.

\textbf{Dual connectivity based densification:} Dual connectivity is an alternative association technique for HUDN. In the dual connectivity mode, a user can connect to the macro BS and a SAP simultaneously, and split the data traffic to these two cells. The link with macro BS is called primary link and the one with SAP is called secondary link.

As shown in Fig.~\ref{centralized}(b), a mobile user can have a primary link and a secondary link. To provide a satisfactory communication quality, these two links need to utilize different channels. Thus, resource allocation in this scenario needs to mitigate interference among different users as well as that between the primary links and the secondary ones.

\subsection{Distributed Network Densification}

For distributed network densification, we select two representative scenarios: D2D based densification and proactive caching based densification.

\begin{figure}[!t]
		\centering
		\includegraphics[width=6.0in]{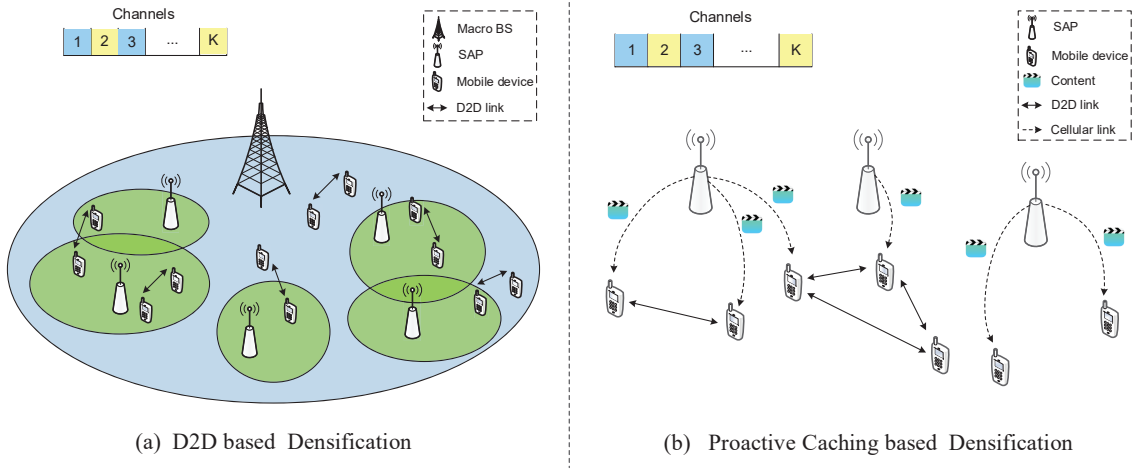}
		\caption{Two representative scenarios for distributed network densification.} \label{distributed}
\end{figure}

\textbf{D2D enabled densification:} D2D communication is a promising technology for 5G cellular networks, where two proximate mobile devices can set up direct communications~\cite{LDZE-2015}. In this way, the network densification can be realized by simply increasing the number of communication links per unit area as illustrated in Fig.~\ref{distributed}(a), where these D2D links can share the same spectrum.
	
Due to the dense deployment of D2D links, resource allocation in this scenario becomes very critical in HUDN. Interference from neighboring D2D users becomes severe. The resource allocation problem in this scenario turns to how to coordinate interference among these D2D links in a distributed way. 
	
\textbf{Proactive caching based network densification:} Proactive caching is to store popular contents predictively at the edge of mobile networks~(e.g., BS, SAP, or mobile devices) to accelerate transmission and reduce the peak traffic loads. By caching the popular contents in the SAPs, users can have better Quality of Experience~(QoE). Besides, it is also possible to reserve the downloaded contents and serve the nearby users by D2D communications. The system model of proactive caching based densification is shown in Fig.~\ref{distributed}(b).
	
Here, we assume that all popular contents have been cached in the SAPs since they are densely deployed. Therefore, the users can either download the desired contents from SAPs or share them from the cached users. The communication procedure is briefly described below:
\begin{itemize}
	\item Mode selection: The users first examine whether the desired contents are cached in SAPs and nearby devices. If the desired ones are only cached in a SAP or a device, the users can only download the contents by cellular or D2D communications, correspondingly. If the desired ones are both cached, the users will select the transmission mode with better SINR.
	
	\item Channel access: Resource management is another important issue. If the network can provide a higher data rate for a larger content, the average QoE will improve. The users will select a channel that satisfies the QoE requirements.
\end{itemize}

The resource allocation in this scenario involves transmission mode selection and channel access.

\section{Hypergraph Theory and Its Applications}
\label{Resource}

In this section, we apply the hypergraph theoretic approaches to solve resource allocation problems in these four aforementioned scenarios. We start with the basic concepts in hypergraph theory. 

\subsection{Basics of Hypergraph Theory}
Hypergraph is a generalized graph, in which any subset of a given set can be an edge, but has fundamental difference from the conventional graph. Here is the definition of hypergraph~\cite{V-2008}.

\begin{figure}[!t]
	\centering
	\includegraphics[width=4.0in]{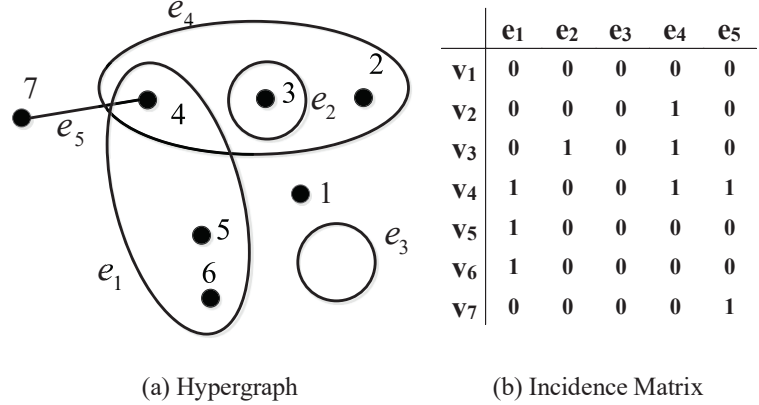}
	\caption{Illustration for a hypergraph and its incidence matrix.} \label{basics}
\end{figure}

\begin{definition}\label{hypergraphdefinition}
	Let $V = \{v_1,v_2,\ldots,v_n\}$ be a finite set, a hypergraph $H$ on $V$ is a family $E = (e_1,e_2,\ldots,e_m)$ of subsets of $V$ such that
	\begin{equation}\label{Hypergraph_definition}
	\bigcup\limits_{i = 1}^m {e_i}  = V.
	\end{equation}
	The elements $v_1,v_2,\ldots,v_n$ of $V$ are vertices of hypergraph $H$, and sets $e_1,e_2,\ldots,e_m$ are the hyperedges of hypergraph $H$.
\end{definition}

As shown in Fig.~\ref{basics}, the hypergraph can be also specified from its incidence matrix. The incidence matrix has one row for each vertex and one column for each hyperedge. If vertex $v_i$ is incident to hyperedge $e_j$, then $(i,j)$-entry in the matrix is 1, otherwise it is 0. In this figure, the hyperedge $e_3$ does not contain any vertex, and thus, in the incidence matrix, all the elements are 0.

In this article, we will introduce three hypergraph based approaches to solve resource allocation problems in HUDN: hypergraph matching~\cite{LHYWH-2016}, hypergraph coloring~\cite{HLH-2016}, and hypergraph game~\cite{YQYYFJ-2017}. Their characteristics are summarized respectively as follows:

\begin{itemize}
	\item \textbf{Hypergraph matching:} Hypergraph matching problem is a multiple dimensional matching where the entities are divided into several disjoint groups with one entity in each group. As each vertex only requires the information from the adjacent vertices in the hypergraph matching, it is a distributed approach.  
	
	\item \textbf{Hypergraph coloring:} Hypergraph coloring, which arises from the map coloring, has been widely used in many-to-one assignment problems, such as scheduling and resource allocation. The hypergraph coloring requires a centralized node that has all the information of the hypergraph and selects a color for each vertex.
	
	\item \textbf{Hypergraph game:} Hypergraph game is also suitable to tackle the many-to-one assignment problems, especially the cases where the utility of each vertex is only related to the vertices in the hyperedges. In contrast to hypergraph coloring, hypergraph game is a distributed method, where each vertex calculates the utilities through information exchange among different vertices and uses the strategy with the highest utility.
\end{itemize}

The main advantage of hypergraph theory is its ability to model the relations among multiple entities compared to other heuristic approaches, such as game and conventional matching theory. In game theory, each player can only calculate its utility for each strategy, and thus, it cannot model the relations among different entities. Conventional matching theory cannot model the relations among multiple entities, either. In conventional matching theory, it is possible for two agents to calculate the utility if they form a matching pair but we cannot effectively evaluate the mutual influence among more than two agents. Here are their differences.

\begin{itemize}
	\item Hypergraph matching is similar to conventional matching theory in the sense that they both want to find a matching among agents. But, they also have some differences. In the hypergraph matching, matching results can only be selected from the hyperedges and the hyperedges are predefined before the matching procedure. However, in the conventional matching theory, the matching result can be any matching in the set of definition. Each agent can select any other agent to form a matching by calculating the utility function in the matching procedure.
	
	\item Overlapping coalition formation game~\cite{TLZW-2016} is similar to the hypergraph game but has different perspectives. In hypergraph game, the topology of hypergraph will not change, that is, the vertices are not allowed to be removed from or added into the hyperedge in the operation. While in the overlapping coalition game, the players will be added to or deleted from the coalitions to improve the overall utilities.
\end{itemize}

\subsection{Centralized Resource Allocation}

Here, we will introduce hypergraph matching and coloring for centralized resource allocation.

\begin{table}
	\centering
	\caption{Hypergraph theoretic approaches for typical scenarios in HUDN} \label{Summary}
	\begin{tabular}{|p{2.5cm}|p{1.5cm}|p{7.0cm}|p{3.0cm}|}
		\hline Typical scenario & Category & Hypergraph model & Solution\\
		\hline \makecell[l]{CRAN based \\ densification} & Centralized & \makecell[l]{Vertices: users, channels and RRHs; \\ Hyperedges: the minimum communication requirement \\ is satisfied;\\ Weight: data rate}& Hypergraph matching\\
		\hline \makecell[l]{Dual connectivity \\ based densification} & Centralized &  \makecell[l]{Vertices: primary and secondary links; \\ Hyperedges: interference relations; \\ Colors: channels} & Hypergraph coloring\\
		\hline \makecell[l]{D2D based \\ densification} & Distributed &  \makecell[l]{Vertices: D2D links; \\ Hyperedges: interference relations; \\ Utility: negative number of adjacent vertices with\\ the same channel} & Hypergraph game\\
		\hline \makecell[l]{Caching based \\ densification} & Distributed &  \makecell[l]{Vertices: users, SAPs and channels; \\ Hyperedges: potential matchings; \\Weight: related to data rate and content length} & Hypergraph matching\\
		\hline
	\end{tabular}
\end{table}

\textbf{Hypergraph matching for CRAN based densification:} In the resource allocation problem for CRAN based densification, we assume that one user can only connect to one RRH and be assigned one channel. Under this assumption, the resource allocation can be formulated as a 3-dimensional hypergraph matching problem\footnote{Although hypergraph matching is a distributed approach, it can be also applied in a centralized scenario. These distributed computations can be done in the centralized node in parallel.}.

In the hypergraph matching model, we formulate the users, channels, and RRHs as three disjoint vertex groups, and the hyperedges indicate the relations among these vertices. The hyperedges are formed only if the quality of the link satisfies the minimum requirement, and its weight is defined as the data rate. In this way, we can search the optimal matching for channel allocation. 
 
\textbf{Hypergraph coloring for dual connectivity based densification:} In the resource allocation with dual connectivity, a mobile user has two links that will utilize two orthogonal channels. Therefore, the aforementioned hypergraph matching approach cannot be used, and the hypergraph coloring is adopted to formulate this problem.

In the hypergraph coloring model, the links are regarded as the vertices, and the interference relation is formulated as a hyperedge. When independent interference or cumulative interference exceeds a predefined threshold, these links form hyperedges. Note that the primary and secondary links of a user are unable to utilize the same channel, thus, they will also form a hyperedge. Besides, we formulate the channels as different colors. Then we can color this hypergraph such that the vertices contained in a hyperedge are not colored uniformly. The coloring result corresponds to channel allocation to these links.

\subsection{Distributed Resource Allocation}

Hypergraph game and matching can be used for distributed resource allocation.

\textbf{Hypergraph game for D2D based densification:} In resource allocation of D2D based densification, interference mainly comes from the nearby D2D links. We assume that each D2D link is aware of the set of interfering users from which cumulative interference will affect the communication quality and can access the information of the selected channel for the nearby D2D links. In this way, each D2D link can select the channel to access in order to minimize the total interference level. Therefore, it is suitable to apply hypergraph game to solve this problem. 

In the hypergraph game model, the D2D links are formulated as the vertices, the interference relations among nearby D2D links are regarded as hyperedges, and the utility of each player~(vertex) over a channel is defined as the negative number of adjacent vertices\footnote{Two vertices are said to be adjacent when they are contained by one hyperedge.} that also access this channel. Each D2D link will first calculate its access probability on each channel, which is proportional to the utility, and then access one channel randomly according to its access probability. Since each D2D link intends to select the least interfered channel, the overall performance improves after the hypergraph game.

\textbf{Hypergraph matching for proactive caching based densification:} In the resource allocation of the proactive caching based densification, we assume that each user only requires one content in a time. Besides, the SAP can provide at most one channel for a communication link. In this setting, the resource allocation problem can be transformed to a 3-dimensional hypergraph matching problem.

Note that the resource allocation involves transmission mode selection and channel access. Therefore, in the hypergraph matching model, the vertex set needs to contain  user, SAP, and channel subsets. It should be mentioned that we add some virtual vertices to the SAP subset in order to represent the D2D transmission mode. In this considered network, the vertices will form a hyperedge when
\begin{enumerate}
	\item The user requires a content that is cached in the SAP. If the content is also cached in the nearby users, the user will connect to the virtual vertices.
	
	\item The SINR over a channel needs to satisfy the transmission requirement of the content.
\end{enumerate} 
Besides, to improve the overall QoE, the shorter content has the higher transmission priority. Therefore, the weight of a hyperedge can be defined as a function that is negatively related to the transmission rate and positively related to the content length. In this way, the hypergraph is constructed, and each user can select an optimal hyperedge independently.  

\section{Hypergraph Matching for CRAN based Network Densification} 
\label{OneExample}

In the previous section, we have briefly discussed hypergraph theory, matching, coloring, game, and their potential applications. In this section, we focus on  hypergraph matching for resource allocation in the CRAN based densification. 

\subsection{Basics of Hypergraph Matching}

There is a special hypergraph called $r$-uniform hypergraph if the set of vertices is partitioned into $r$ disjoint subsets such that every hyperedge only contains exactly one vertex in each subset. Therefore, the $r$-dimensional hypergraph matching problem is to find a hyperedge subset of a $r$-uniform weighted hypergraph with the maximum total weight.

However, the $r$-uniform hypergraph matching problem is NP-hard for $r \geq 3$. Thus, the local search algorithm \cite{LHYWH-2016} is utilized to solve the hypergraph matching problem efficiently. To facilitate understanding of the local search algorithm, we first introduce an important concept of $k$-claw.

\begin{definition}\label{k-claw}
Given hypergraph $H=(V,E)$, the representative graph of $H$ is the graph $L(H) = (V',E')$ where every hyperedge in $E$ is represented by a vertex in $V'$, and the intersection of two hyperedges corresponds to an edge in $E'$. In the representative graph, a $k$-claw $C_k$ is defined as a subgraph of $L(H)$ whose center vertex connects to $k$ independent vertices~(talons).
\end{definition}

The local search begins with an initial matching that can be obtained by some greedy algorithms. Given a hyperedge (a center vertex in the representative graph), we search for a $k$-claw that can improve the overall performance for all $2 \leq k \leq r$. If there exists such a $k$-claw, we add the containing talons to the matching and remove all hyperedges that intersect with them from the matching. The above process will be repeated until all $k$-claws are examined. Therefore, this is a greedy method that can obtain a suboptimal solution. 

In the following, we will provide an example to show how the hypergraph matching can be used in resource allocation for CRAN based densification.

\subsection{Resource Allocation}

As shown in Fig.~\ref{centralized}(a), the network consists of a BBU pool with computation functions and RRHs to provide coverage. The users need to associate with one RRH first and the BBU pool assigns one channel to each link for data transmission. Thus, this user association and channel assignment problem can be formulated as a hypergraph matching. The objective is to maximize the data sum-rate by optimizing the matching variables user-RRH-channel. 

We assume that there exist $M$ users, $K$ channels, and $N$ RRHs in this system. In the hypergraph matching model, the users, channels, and RRHs correspond to three disjoint vertex subsets, and a hyperedge implies that the contained vertices are potential to be a matching. Due to interference, it is impossible to define the weight in each hyperedge, and thus, we propose an iterative hypergraph matching algorithm with a complexity of $O(MNK(M+N+K)^6)$ to tackle this problem sufficiently. In each iteration, each hyperedge will update its weight and the selected hyperedges cannot have a common channel vertex, i.e., a channel can be matched with one user.

The matching algorithm can be briefly described below:
\begin{itemize}
	\item \textbf{Step 1:} Construct the hypergraph and initialize the weights of hyperedges. For instance, we can define the data rate when the user-RRH link utilizes the channel independently as the initial weight for a hyperedge. 
	
	\item \textbf{Step 2:} Apply the aforementioned local search algorithm to find a matching that does not have a common channel vertex. Besides, delete those hyperedges that have the same users as that in this matching from the hypergraph.
	
	\item \textbf{Step 3:} Update the weights of hyperedges according to the current matching. The weight of each hyperedge is defined as the data rate improvement when this hyperedge is added to the current matching. If the data rate improvement is negative (the data rate is reduced), this hyperedge is removed from the hypergraph.
	
	\item \textbf{Step 4:} Repeat Steps 2 and 3 until the hypergraph has an empty hyperedge set.
\end{itemize}

\begin{figure}[!t]
 	\centering
 	\subfigure[]
 	{
 	\begin{minipage}{3.0in}
 	\centering
 	\includegraphics[width=3.0in]{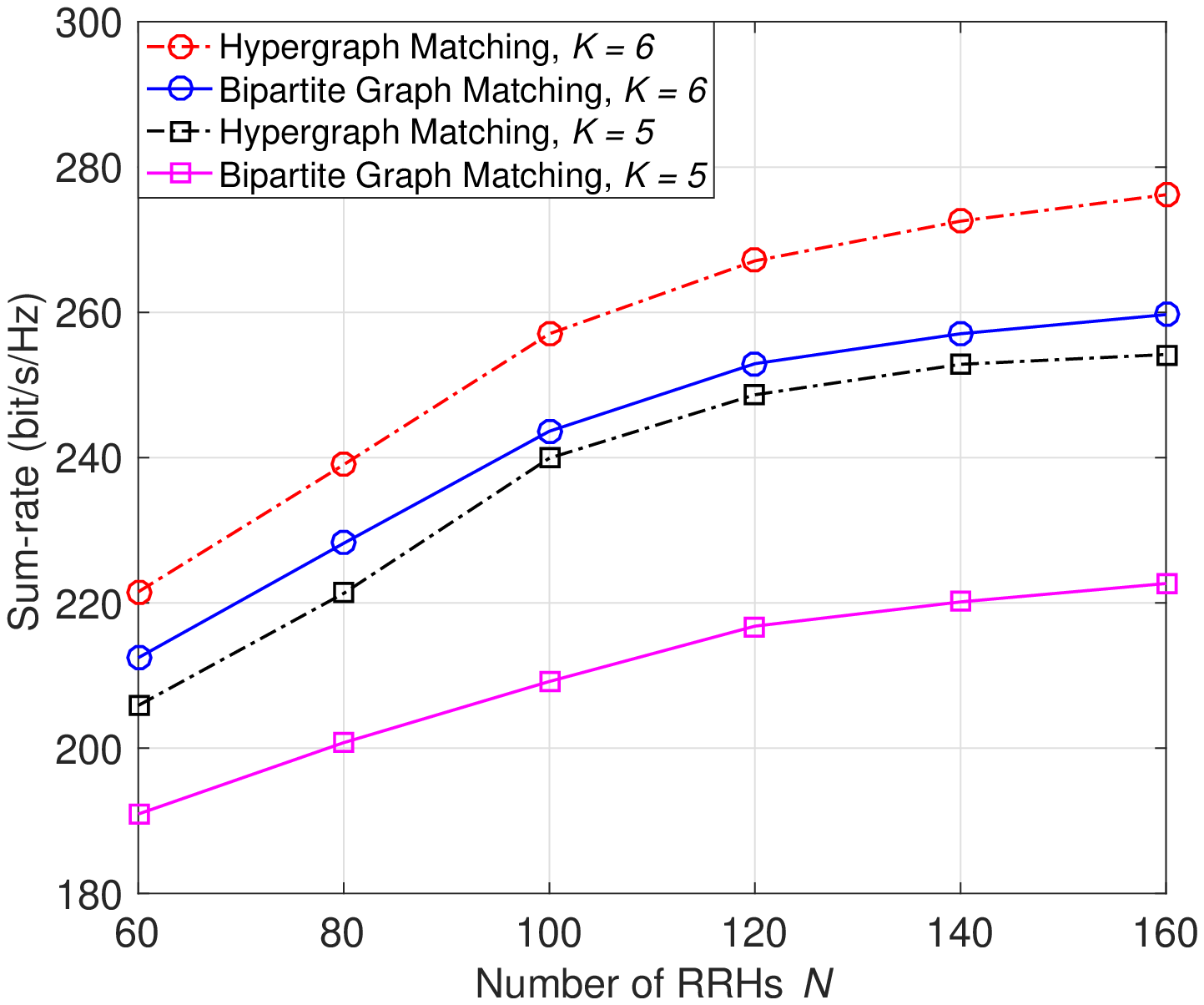}\label{sumrate}	
 	\end{minipage}
 	}
	 \subfigure[]
	 {
	 	\begin{minipage}{3.0in}
	 		\centering
	 		\includegraphics[width=3.0in]{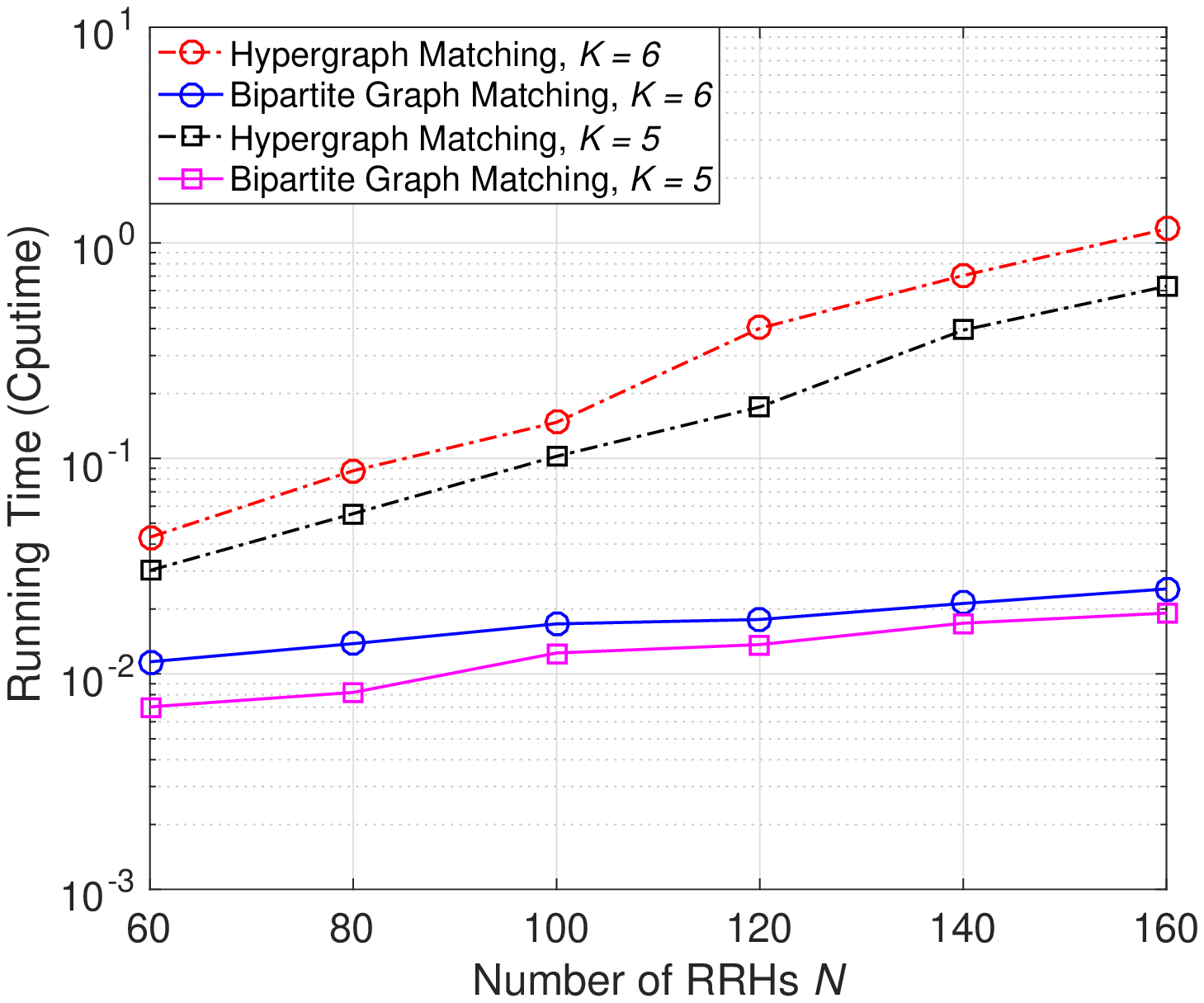}\label{complexity}	
	 	\end{minipage}
	 }
 	\caption{System performance: (a) Sum-rate versus Number of RRHs $N$ with $M = 20$; (b) Running time versus Number of RRHs $N$ in CRAN based densification with $M = 20$.} \label{Performance}
 \end{figure}

In Fig.~\ref{Performance}, we compare the hypergraph matching algorithm with the bipartite graph matching algorithm\footnote{In bipartite graph matching algorithm, the users always associate with the nearest RRH, and channel assignment is performed by iterative Hungarian Algorithm~\cite{HTLZ-2013}.} in terms of the sum-rate as well as the computational complexity. From the figure, the hypergraph matching algorithm outperforms the bipartite graph matching algorithm for different values of $N$ though it costs more computational time. This can help the operators to determine the number of deployed RRHs according to the computational resources in the BBU pool. 

\section{Conclusions and Future Outlook}
\label{Conclusion}
In this article, hypergraph theory has been introduced to model the relation among different elements in HUDN. Due to the dense deployment of SAPs, the interference from the neighboring cells becomes severe and resource allocation also involves user association. The hypergraph theoretic models can consider cumulative interference and user association simultaneously and are useful for designing efficient radio resource allocation algorithms in these scenarios. In centralized network densification, hypergraph matching and coloring are respectively used to solve resource allocation for CRAN and dual connectivity based densification. In distributed network densification, a hypergraph game is applied to perform channel access in a distributed manner and the hypergraph matching can address the resource allocation in proactive caching based densification. Besides, we have presented a hypergraph matching model for the resource allocation problem of CRAN based densification in detail.

As an effective approach modeling the complicated relations among multiple parties, hypergraph theory shows its power in the resource allocation for HUDN. There still exist many open issues in this field, including the two examples listed below.

\textbf{Cognitive radio based densification:} To further improve the spectrum efficiency, it is natural to extend the HUDN to the unlicensed spectrum where the users in SCs can only utilize the unlicensed spectrum~\cite{FLHJ-2015}. Different from the aforementioned channel access methods, the users access the unlicensed spectrum in a cognitive way, i.e., the secondary users (e.g., machine-to-machine link) need to sense the spectrum hole before transmission. In this scheme, the hypergraph game can also be applied to solve the spectrum sharing problem by formulating the relations between the primary and the secondary users as hyperedges.

\textbf{Network virtualization based densification:} Wireless network virtualization is a promising technology to accommodate the significant growth in wireless traffic, in which physical resources are abstracted and shared among different parties \cite{CFHZ-2016}. With virtualization, the network infrastructure is separated from the services, thereby providing convenience for the operation in HUDN. In such a virtual network, the resource allocation schemes not only need to decide how to distribute the content in the virtual networks, but also how to map the virtual networks with the physical ones. This kind of relations among multiple entities can also be modeled by hypergraph theory. By means of hypergraph matching, the joint content distribution and physical resource mapping problem can be solved efficiently.


\section*{Acknowledgements}%

This work was supported by National Science and Technology Major Project of China under grant number 2016ZX03001017 and the National Nature Science Foundation of China under grant number 61625101 and 61511130085.

\begin{IEEEbiography}
	{Hongliang Zhang} [S'15] (hongliang.zhang@pku.edu.cn) received the B.~S.~degree in Electronic Engineering from Peking University in 2014. He is currently pursuing his PhD's degree in Peking University. His current research interest includes wireless communications, hypergraph theory, and optimization theory.
\end{IEEEbiography}

\begin{IEEEbiography}
	{Lingyang Song} [S'03, M'06, SM'12] (lingyang.song@pku.edu.cn) is a Professor in School of Electronics Engineering and Computer Science, Peking University. His research interests include MIMO, cognitive and cooperative communications, physical layer security, and wireless ad hoc/sensor networks. He is the recipient of 2012 IEEE Asia Pacific Young Researcher Award and 2016 IEEE ComSoc Leonard G. Abraham Prize. He is an editor for \emph{IEEE Transactions on Wireless Communications}. He has been an IEEE Distinguished Lecturer since 2015.
\end{IEEEbiography}

\begin{IEEEbiography}
	{Yonghui Li} [M'04, SM'09] (yonghui.li@sydney.edu.au) received his PhD degree in 2002 from Beijing University of Aeronautics and Astronautics. He is a Professor at University of Sydney. He is the recipient of 2008 Australian Queen Elizabeth II Fellowship and 2012 Australian Future Fellowship.  His current research interests focus on MIMO, millimeter wave communications, machine to machine communications, coding and cooperative communications. He is an editor for \emph{IEEE Transactions on Communications} and \emph{IEEE Transactions on Vehicular Technology}. 
\end{IEEEbiography}

\begin{IEEEbiography}
	{Geoffrey Ye Li} [S'93, M'95, SM'97, F'06] (liye@ece.gatech.edu) is a Professor with Georgia Tech. His research is in signal processing and machine learning for wireless communications. He has published around 400 articles and listed as a Highly-Cited Researcher by Thomson Reuters. He won 2010 Stephen O. Rice Prize Paper Award and 2017 Award for Advances in Communication from IEEE ComSoc, 2013 James Evans Avant Garde Award and 2014 Jack Neubauer Memorial Award from IEEE VTS.
\end{IEEEbiography}


\begin{thebibliography}{20}

\bibitem{Cisco-2015}
Cisco, ``Cisco visual network index: global mobile traffic forecast update 2014-2019 white paper," San Jose, CA, USA, May 2015.

\bibitem{NJDRDARCS-2014}
N.~Bhushan \emph{et al.}, ``Network densification: The dominant theme for wireless evolution into 5G," \emph{IEEE Commun. Mag.}, vol.~52, no.~2, Feb.~2014, pp.~82-89.

\bibitem{MWA-2016}
M.~Kamel, W.~Hamouda, and A.~Youssef, ``Ultra-dense networks: a survey," \emph{IEEE Commun. Surveys \& Tutorials}, vol.~18, no.~4, May 2016, pp.~2522-2545.

\bibitem{CKHPSETB-2016}
C.~Rosa \emph{et al.}, ``Dual connectivity for LTE small cell evolution: functionality and performance aspects," \emph{IEEE Commun. Mag.}, vol.~54, no.~6, Jun.~2016, pp.~137-143.

\bibitem{HTLZ-2013} 
H.~Zhang \emph{et al.}, ``Graph-based resource allocation for D2D communications underlaying cellular networks," in \emph{Proc. IEEE/CIC ICCC}, Xi'an, P.R. China, Aug.~2013, pp.~187-192. 

\bibitem{HLH-2016}
H.~Zhang, L.~Song, and H.~Zhu ``Radio resource allocation for device-to-device underlay communication using hypergraph theory," \emph{IEEE Trans. Wireless Commun.}, vol.~15, no.~7, Jul.~2016, pp.~4852-4861.

\bibitem{LHYWH-2016}
L.~Wang \emph{et al.}, ``Hypergraph-based wireless distributed storage optimization for cellular D2D underlays," \emph{IEEE JSAC}, vol.~34, no.~10, Oct.~2016, pp.~2650-2666.

\bibitem{RSYJ-2013}
R.~Heath \emph{et al.}, ``A current perspective on distributed antenna systems for downlink of cellular systems," \emph{IEEE Commun. Mag.}, vol.~51, no.~4, Apr.~2013, pp.~161-167.

\bibitem{AHYLGML-2014}
A.~Checko \emph{et al.}, ``Cloud ran for mobile networks - technology overview," \emph{IEEE Commun. Surveys \& Tutorials}, vol.~17, no.~1, Sep.~2014, pp.~405-426.

\bibitem{LDZE-2015}
L.~Song \emph{et al.}, \emph{Wireless device-to-device communications and networks}, Cambridge University Press, UK, 2015.

\bibitem{V-2008}
V.~I.~Voloshin, \emph{Introduction to graph and hypergraph theory}. Nova Science Publishers, New York City, NY, 2008.



\bibitem{YQYYFJ-2017}
Y.~Sun \emph{et al.}, ``Distributed channel access for device-to-device communications: a hypergraph-based learning solution," \emph{IEEE Commun. Letters}, vol.~21, no.~1, Jan.~2017, pp.~180-183.

\bibitem{TLZW-2016}
T.~Wang \emph{et al.}, ``Overlapping coalition formation games for emerging communication networks," \emph{IEEE Network}, vol.~30, no.~5, Sep.~2016, pp.~46-53.

\bibitem{FLHJ-2015}
F.-H.~Tseng \emph{et al.}, ``Ultra-dense small cell planning using cognitive radio network toward 5G," \emph{IEEE Wireless Commun.}, vol.~22, no.~6, Dec.~2015, pp.~76-83.

\bibitem{CFHZ-2016}
C.~Liang \emph{et al.}, ``Virtual resource allocation in information-centric wireless networks with virtualization," \emph{IEEE Trans.  Vehic. Tech.}, vol.~65, no.~12, Dec.~2016, pp.~9902-9914.

\end{thebibliography}
\end{document}